%% file: main.tex
\documentclass[conference,letterpaper,10pt]{IEEEtran}

\newcommand{\revision}[1]{\textcolor{red}{\textbf{#1}}}
\renewcommand{\revision}[1]{{#1}}

\usepackage{adjustbox}
\usepackage{array}

\newcolumntype{R}[2]{%
    >{\adjustbox{angle=#1,lap=\width-(#2)}\bgroup}%
    l%
    <{\egroup}%
}
\newcommand*\rot{\multicolumn{1}{R{90}{1em}}}

\usepackage{tikz}
\newcolumntype{C}{@{\extracolsep{.75cm}}c@{\extracolsep{0pt}}}%

\usepackage{wasysym}

\usepackage{multicol}

\usepackage[font=small,labelfont=bf]{caption}

\usepackage{microtype}
\everypar{\looseness=-1}

\usepackage{multirow}
\usepackage{subcaption}
\usepackage{graphics}
\graphicspath{ {images/} }
\usepackage{mwe}
\usepackage{array}

\usepackage{comment}

\usepackage{algorithm}
\usepackage{algorithmic}

\usepackage{float}

\usepackage{arydshln}

\makeatletter
\newcommand{\rmnum}[1]{\romannumeral #1}
\newcommand{\Rmnum}[1]{\expandafter\@slowromancap\romannumeral #1@}
\makeatother

\usepackage{array,booktabs}
\newcolumntype{L}{@{}>{\kern\tabcolsep}l<{\kern\tabcolsep}}
\usepackage{colortbl}
\usepackage{xcolor}

\usepackage{multirow}
\usepackage{bigstrut}

\makeatletter 
\newcommand\semiHuge{\@setfontsize\semiHuge{22.72}{27.38}}
\makeatother

\begin{document}
%
\title{\semiHuge WACA: Wearable-Assisted Continuous Authentication}


\author{\IEEEauthorblockN{Abbas Acar\IEEEauthorrefmark{1}, Hidayet Aksu\IEEEauthorrefmark{1}, A. Selcuk Uluagac\IEEEauthorrefmark{1}, and Kemal Akkaya\IEEEauthorrefmark{2}}
\IEEEauthorblockA{ \IEEEauthorrefmark{1}Cyber-Physical Systems Security Lab (CSL) \\ \IEEEauthorrefmark{2}Advanced Wireless and Security Lab \\
Department of Electrical and Computer Engineering \\
Florida International University \\
\{aacar001,haksu,suluagac,kakkaya\}@fiu.edu}

}




\maketitle

\begin{abstract}
One-time login process in conventional authentication systems does not guarantee that the identified user is the actual user throughout the session.
\revision{However}, it is necessary to re-verify the user identity periodically throughout a login session without reducing the user convenience. \textit{Continuous authentication}
can address this issue. 
\revision{However, existing methods are either not reliable or not usable.}
In this paper, we introduce a usable and reliable method called 
\textit{Wearable-Assisted Continuous Authentication} (WACA).
WACA 
relies on the \textit{sensor-based keystroke dynamics}, \revision{where} the authentication data is acquired through the built-in sensors of a wearable \revision{(e.g., smartwatch)} while the user is typing. 
We implemented the WACA framework and evaluated its performance on real devices with real users. 
The empirical evaluation of WACA reveals that WACA is feasible and its error rate is as low as \emph{$1\%$} with \emph{$30$} seconds of processing time and $2-3\%$ for \emph{$20$} seconds. The computational overhead is minimal. Furthermore, we tested WACA against different attack scenarios. WACA is capable of identifying insider threats with very high accuracy (\emph{$99.2\%$}) and also robust against powerful adversaries such as imitation and statistical attackers. 

\end{abstract}

\IEEEpeerreviewmaketitle

\section{Introduction}\label{sec:intro}

%

The majority of the current user authentication methods rely on password authentication. However, password authentication methods are subject to many security drawbacks~\cite{bonneau2015passwords,grosse2013authentication,securityrisksurvey2016}. Many practical attacks have been demonstrated that the passwords can be either stolen or bypassed~\cite{pass1,pass2,theofanos2016secure}. 
To mitigate these threats, Multi-Factor Authentication (MFA) methods were proposed~\cite{sabzevar2008universal,wu2004secure}. In MFA, the user credentials are checked from two or more independent sources and even if the attacker steals one factor, it would still have to overcome the burden of other factors. MFA is indeed standardized and recommended by some payment and government organizations~\cite{nist,dss}.

Whether it is one-factor or MFA, a one-time login process does not guarantee that the identified user is the real user throughout the login session. Even if it is an insider who has been authorized once, a forever access is provided in most cases not to interrupt the current user. 
Hence, an authentication which re-verifies the user periodically without breaking the continuity of the session is vital~\cite{darpa,google}. 
Moreover, users may share their passwords with family members, friends,  colleagues~\cite{securityrisksurvey2016}, or an already-authenticated user may walk away without locking his/her computing platform (e.g., laptop)  for a short time or may intentionally hand it to a non-authenticated co-worker  trusting that s/he will not perpetrate anything 
malicious, or disgruntled worker or a malicious former employer may want to use his/her former privileges. In all these cases, as long as the original login session is actively used, there is no mechanism to verify that the initial authenticated user is still the user in control of the  computing terminal.

\textit{Continuous Authentication} (CA)\footnote{CA is also sometimes called Active or Implicit Authentication in the literature~\cite{darpa}.} is a good mechanism to re-verify a user's identity periodically throughout a login session. However, this may pose inconvenience to the users. Currently, the most common method used to verify the user periodically depends on session time-outs. In session time-outs, if the time period is kept too short, the user's convenience will be reduced due to frequent interruptions of the session for authentication. On the other hand, if the time period is set too long, in the case of a breach, the attacker would have more time on the victim's system. Moreover, the insider threat detection is also an important functionality that needs to be considered in continuous authentication systems as a potential attacker is likely to be an insider~\cite{eberz2015preventing}. Indeed, the usability of CA systems can be increased by exploiting off-the-shelf wearable devices as they offer many useful functionalities like controlling the environment, keeping track of the daily activities to their users through their built-in sensors (e.g., motion sensors, heart rate monitoring sensor, and GPS)~\cite{watch,forbes}. These sensors can play a key role to increase the usability in such a security context as well. 

In this work, we introduce a \textit{Wearable-Assisted Continuous Authentication} framework called WACA, where a wearable device (e.g., smartwatch) is used to authenticate a computer user continuously utilizing the motion sensors of the smartwatch. WACA uses \textit{sensor-based keystroke dynamics}, where the typing rhythm of the user is captured by the motion sensors of the smartwatch worn by the user. In essence, keystroke dynamics is one of the behavioral biometrics that characterizes the users according to their typing pattern. Most conventional keystroke-based authentication schemes~\cite{teh2013survey} have used \textit{dwell-time} and \textit{flight-time} as unique features of the users. These features are directly obtained by logging the timing between successive keystrokes. However, in WACA, the feature set is richer and more flexible since 6-axes motion sensor data can provide not only timing information, but also the key-pressing pressure, hand rotation, and hand displacement, etc. 
Our feature set consists of 14 different sensory features from both time and frequency domains. These features are applied to 6-axes motion sensor data, obtaining 84 features in total. Finally, different distance measures are used to compare the registered and the unknown profile of the user as it was shown that they performed well in similar contexts~\cite{killourhy2009comparing,serwadda2013verifiers}.

We tested the performance and efficiency of WACA with more than thirty real users and data collected from them. We specifically evaluated WACA in terms of three metrics: (\rmnum{1}) \textit{How accurately can it differentiate between genuine and impostor users?} (\rmnum{2}) \textit{How fast can it  detect an impostor?} (\rmnum{3}) \textit{How accurately can it identify an insider?} Moreover, we also evaluated the robustness of our proposed method against powerful attacks, including, \textit{imitation}~\cite{tey2013can,huhta2015pitfalls},  \textit{statistical}~\cite{serwadda2013examining,stanciu2016effectiveness}, and insider attacks. 


\noindent \textbf{Contributions:} The main contributions of this work are summarized as follows:
\begin{itemize}
    \item We propose a comprehensive sensor-based wearable-assisted continuous authentication framework for computing platforms, terminals (e.g., laptops, computers) with a smartwatch. 
    We believe that this work has practical and far-reaching implications for the future of the usable authentication field. 
    \item \revision{We propose a new variant of keystroke dynamics, called \emph{sensor-based keystroke dynamics}. We show that \emph{sensor-based keystroke dynamics} can be uniquely utilized to authenticate and identify the users with extesive evaluation.} 
    \item \revision{We also show that, contrary to recent plausible attacks on classical keyboard-only keystroke dynamics~\cite{tey2013can,serwadda2013examining}, this new authentication factor is robust against powerful attacks, including imitation, statistical attacks, and allows insider threat identification.}
    
    \item 
    \revision{For this purpose, we developed three generic attacking scenarios that can also be utilized by other future continuous authentication studies.}
    
    \item \revision{We conducted an extensive evaluation of the proposed method with real devices and real user data using a rich set of distance measuring techniques for the authentication and Multilayer Perceptron (MLP) 
    algorithm for the identification.}
   \end{itemize}

\noindent \textbf{Organization:} The reminder of this paper is structured as follows: In Section~\ref{sec:prelim}, we explain the foundation for the overall idea. Then, we introduce our system model in Section~\ref{sec:adv}. The overall architecture of WACA is detailed in Section~\ref{sec:arc}. Section~\ref{sec:eval} presents the performance, efficiency, and robustness evaluation of the WACA framework. It also evaluates how WACA defends against powerful adversaries. Section~\ref{sec:disc} reports the discussion of the challenges that can be faced in WACA and ways to overcome those challenges. In Section~\ref{sec:related}, we explain the related work and we make the comparative evaluation of WACA with its alternatives for continuous authentication. Finally, in Section~\ref{sec:conc}, we conclude the paper.

\input{design_rationale}
\input{system_model}
\input{waca_architecture}
\input{evaluation}

\input{advanced_attacks}
\input{discussion}

\input{limitations}
\input{related_work}

\input{conclusion}


\section*{Acknowledgment}

This work is partially supported by US National Science
Foundation (NSF) under the grant numbers NSF-CNS-1718116 and NSF-CAREER-CNS-1453647. The statements made herein are solely the responsibility of the authors.



%

\bibliographystyle{IEEEtran}
\bibliography{references}  

\appendix

\input{appendix}

\end{document}

%% file: design_rationale.tex
\section{Design Rationale: Why Should it Work?} \label{sec:prelim}

In this section, we show 
how motion sensors of a smartwatch are simply impacted when typing on a keyboard and see if the data can be really used to identify users. Particularly, we analyze a case that a user wears a smartwatch and types on a qwerty-type built-in keyboard of a computer. Our goal is to collect keystroke information from the built-in motion sensors (i.e., accelerometer and gyroscope) of the smartwatch during the typing activity. To collect smartwatch sensor data, we developed an Android Wear app that records the raw sensor readings from the motion sensors.

\begin{figure}
\raggedright
  \begin{subfigure}{0.22\textwidth}
    \includegraphics[scale=0.7]{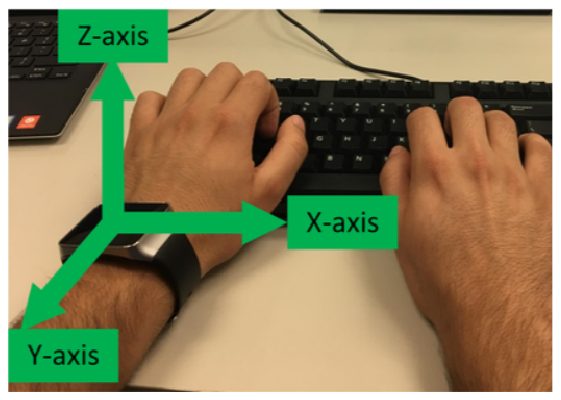}
   \caption{\label{coordinates}}
 \end{subfigure}
 \hspace{10pt}
  \begin{subfigure}{0.22\textwidth}
    \includegraphics[scale=0.7]{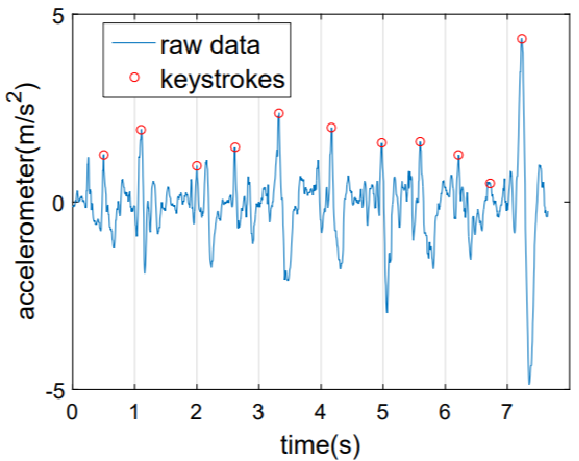}
       \caption{\label{sample-acc-data}}
  \end{subfigure}
  \caption{(a) The reference coordinate system for accelerometer and gyroscope sensors. (b) A sample raw data collected from the accelerometer of the smartwatch and keystrokes detected by using peak detection methods while typing the word "smartwatch". \label{design} }
  \vspace{1pt}
\end{figure}

In our experiments, we used linear acceleration composite sensor data, which combines the data of accelerometer and gyroscope to exclude the effect of gravity\footnote{For brevity we use acceleration to refer to the linear acceleration.}. Note that the accelerometer and gyroscope sensors provide three dimensional sensor data, where the reference coordinate system associated with the sensors are illustrated in Figure~\ref{design}a. As the z-axis of the accelerometer sensor is directly affected by the key up-down movements of a user while typing, the most significant changes are observed in the z-axis. Therefore, the z-axis of the data provides the best information for keystroke features such as holding time, pressing pressure, etc. Moreover, another observation is that even if the device is placed flat on a desk, the sensors generate a certain level of noise, which needs to be removed by filtering as explained later.

\begin{figure*}[t!]
    \centering
\begin{minipage}{0.48\textwidth}
    \centering
    \includegraphics[width=\textwidth]{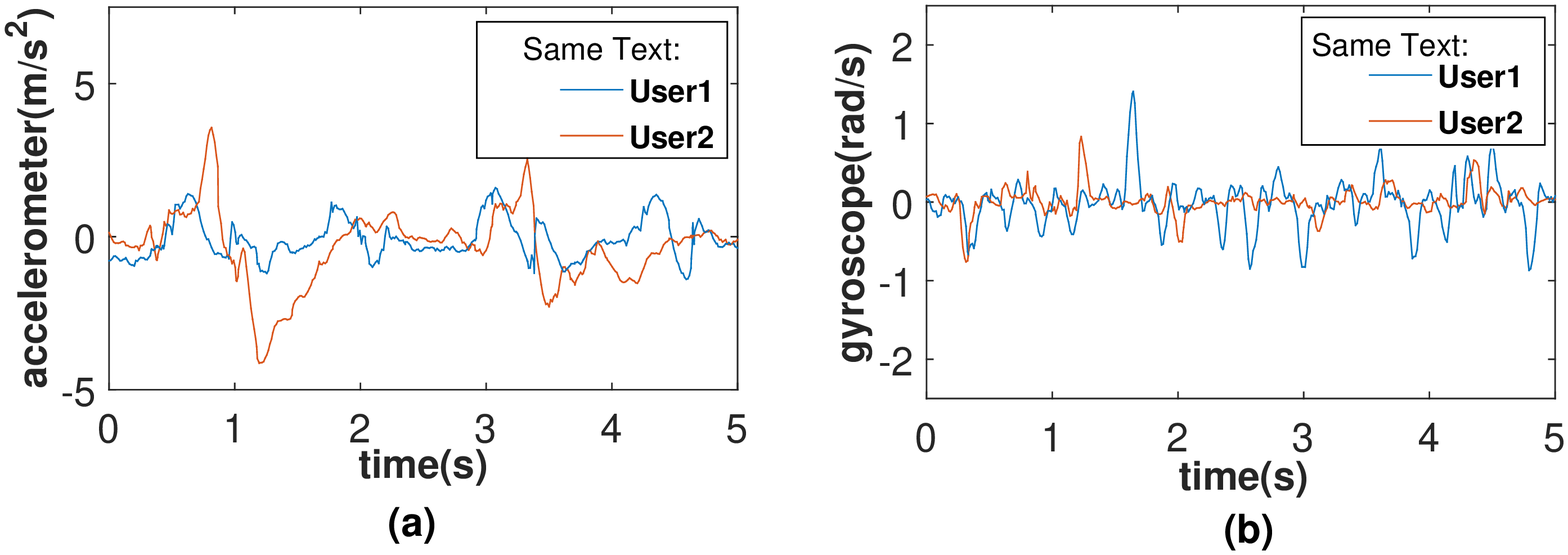}
    \caption{Comparison of two different users' (a) accelerometer (b) gyroscope readings while typing the same text. \label{observe1}}
\end{minipage}\hfill
\begin{minipage}{0.48\textwidth}
    \centering
    \includegraphics[width=\textwidth]{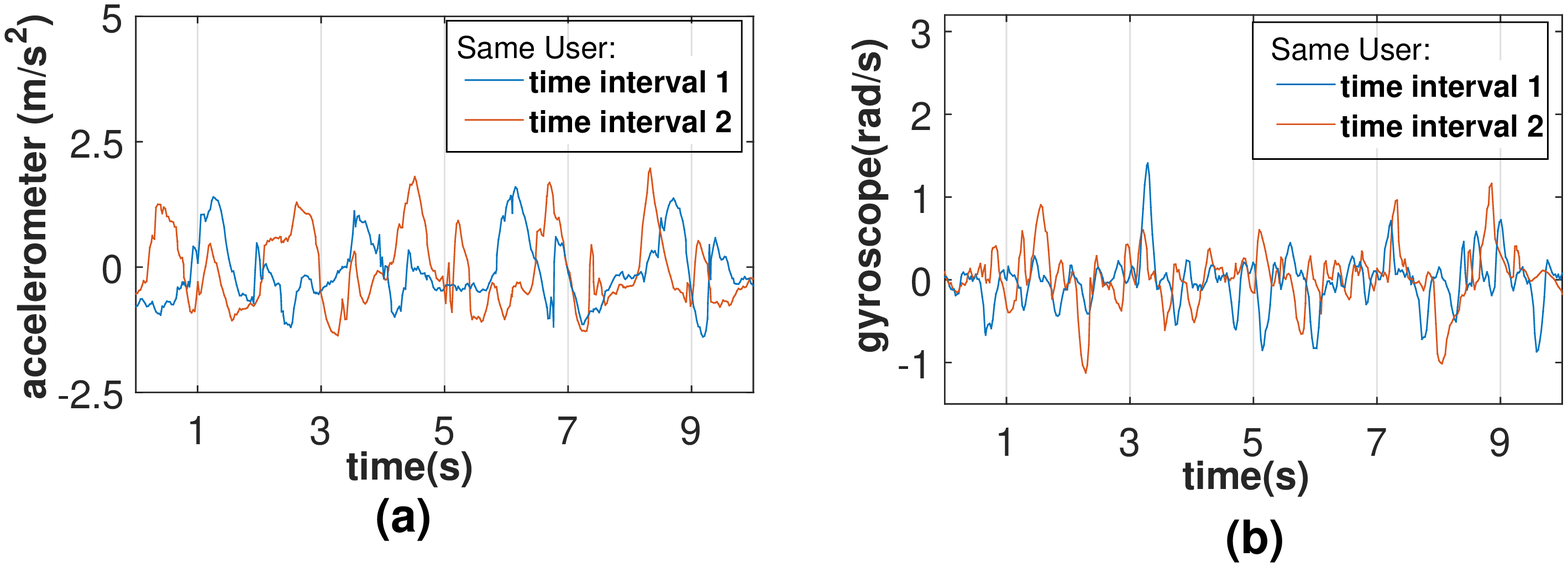}
    \caption{Comparison of the same user's sensor data over two different time intervals with (a) accelerometer, (b) gyroscope. \label{observe2}}
\end{minipage}    
\end{figure*}

Sample data in Figure~\ref{design}b was acquired from the z-axis of the accelerometer while typing the word ``smartwatch". It can be seen how the value of the accelerometer makes peak points. As the acceleration through the gravity corresponds to the going down of the accelerometer, the peak points in the figure correspond to the keystrokes in the typing activity. While the amplitude of the peak is related to how strong the key press is, the width of the peaks is associated with how long the key is pressed. These are simple statistics that can be used to identify the users. Comprehensive list of all the features used in WACA is given in Section~\ref{sec:arc} and will be analyzed further in detail.
\par Moreover, we conducted two more simple experiments using the accelerometer and gyroscope data on the smartwatch and we made the following two observations:

\begin{itemize}
    \item \textit{Hypothesis 1: Different users exhibit different patterns even if they type the same text.}
\end{itemize}

In this experiment, we compared the data collected from two different users while typing the same text. Figure~\ref{observe1} presents the sensor data of the two users' accelerometer and gyroscope data for a given time interval. The distribution of the accelerometer data in Figure~\ref{observe1}a shows clear differences such as the magnitude of peaks, inter-arrival time of peak points, width of peaks, etc. On the other hand, the gyroscope sensor measures the rotation of the watch.  As seen in Figure~\ref{observe1}b, the number of peaks or the magnitude of the peaks are different for different users; so these features are viable candidates to recognize different users.

\begin{itemize}
    \item \textit{Hypothesis 2: Same user follows similar patterns over different time intervals even while typing different texts.}
\end{itemize}

In the second experiment, the data was collected from the same user over two different time intervals corresponding to the different texts and the plots are given in Figure~\ref{observe2}. As seen in Figure~\ref{observe2}a, the amplitudes and widths of the peaks are similar in magnitude, but with a phase shift, meaning leading or lagging. On the other hand, the same leading or lagging of similar shapes can also be seen in the gyroscope data in Figure~\ref{observe2}b.


These two hypotheses justify the rationale that keystroke dynamics obtained from smartwatch' accelerometer and gyroscope sensors can differentiate different users as classical keystroke dynamics and same users can be detected over different times even while typing different texts. Although these are just preliminary observations, our framework are further tested and evaluated with extensive experiments using real user data in Section~\ref{sec:eval}.

%% file: system_model.tex
\section{Design Goals, Assumptions, and Adversary Model}\label{sec:adv}
In this section, we explain our design goals, assumptions, and the adversary model.

\noindent \textbf{Design Goals:} In WACA, our design goals is similar to the ones suggested
in~\cite{jain2006biometrics}: Our system aims to be \textit{universal} (i.e., the biometric features exist for everyone), \textit{unique} (the features are specific for everyone), \textit{permanent} (the biometric features  always exist), \textit{unobtrusive} (the system works with minimal burden), \textit{transparent} (the system works without interrupting the user), \textit{continuous} (the system should provide continuous user data), and \textit{accurate} (the system works with low error rate). WACA achieves the first five goals by its design and the accuracy is tested in Section~\ref{sec:eval}. 

\noindent \textbf{Assumptions:} For WACA, the following assumptions are made:
\begin{itemize}
    \item We assume that the user wears a smartwatch, which is equipped with motion sensors and either Bluetooth or WiFi. We also assume that an app to collect the motion data is already installed on the smartwatch and it is paired with the computer that will be authenticated. For this, we built a custom Android Wear app to collect and process the smartwatch sensor data.
    \item We assume that by pairing devices, a secure communication channel is already established between the computer and smartwatch as well as between the computer and the remote or local authentication server. This secure communication channel should keep the sensor data secure in both transition and at rest. Note that  pairing is only needed for creating a secure channel between the smartwatch and terminal. A standard encryption algorithm using Bluetooth can be utilized for this purpose. 
    \item The WACA framework acts like as a complementary to the first-factor in the authentication and it has the flexibility to choose the first factor, but we assume the system has a first authentication factor. The first factor could be one of the password-, token-, or biometric-based systems.
\end{itemize}

\noindent \textbf{Adversary Model:} In this paper, the primarily considered adversary model is an attacker who somehow bypassed the first factor (e.g., password, token) of the authentication system and it has a physical access to the computing terminal. Particularly, we consider an authentication setting, where users are given passwords to use some computers and for some computers, only high level supervisors are allowed to use. In this type of environment, the attacker is likely to be an insider or co-worker, but it can also be an outsider, just passing by the victim's computer. For example, attacker's goals can include, but not limited to, trying to get some important information from the victim's computer, taking action on behalf of the victim, or trying to get access to the assets that s/he does not have permission (i.e., privilege abuse). 

\begin{figure*}
  \centering
  \resizebox{.8\textwidth}{!}{
  \framebox{
  
  \centering
    \includegraphics[clip, trim=0cm 0cm 0cm 0cm, width=\textwidth]{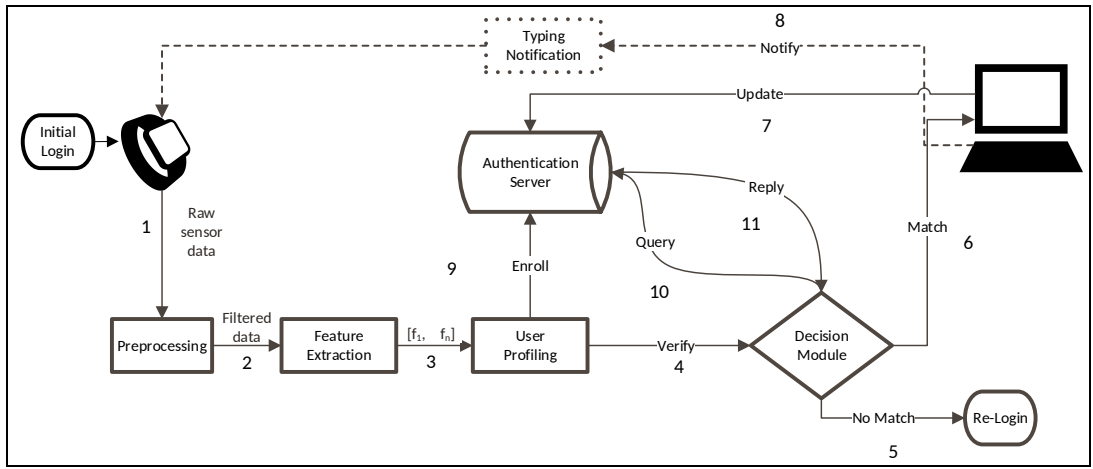}}}
    \caption{WACA framework architecture and key components}
      \label{overall}
\end{figure*}

More specifically, we consider the following attack scenarios:
\begin{itemize}
    \item \textit{Attack Scenario 1:} The victim is one of the employers and forgets to lock his computer and \textit{an outsider} (e.g., a mail courier) who is just passing through the office tries to get access to the victim's computer. 
    Since WACA assume  all the legitimate users' including insiders' smartwatches are paired with the computing terminals, the system will easily reject the outsiders as no input from the smartwatch is provided. Even if the attacker has somehow obtained the smartwatch (i.e., the victim can forget it on the desk), the attacker can not provide the victim' typing pattern (as a behavioral biometric).
    \item \textit{Attack Scenario 2:} We consider the attacker is either an innocent co-worker or a malicious insider and thereby the attacker also has a registered smartwatch, but its typing profile is registered together with its own username. This type of attacker tries to get access to the system's assets that s/he does not have permission (i.e., privilege abuse). In this scenario, the attacker watches its victim (e.g., supervisor) for a suitable timing that its victim leaves the computer unlocked for some time to go to lunch or to get coffee etc. (aka \textit{lunchtime attack}~\cite{eberz2015preventing}). 
    The attacker can either try to bypass the system via providing data from his smartwatch or can try to use the victim's smartwatch somehow obtained (e.g., can steal it or victim can leave it behind).
    \item \textit{More Powerful Adversaries:} Furthermore, a powerful adversary can be aware of WACA and try to defeat it using special tools and skills by \textit{imitating} legitimate users~\cite{tey2013can,huhta2015pitfalls} or \textit{launching statistical} attacks~\cite{serwadda2013examining,stanciu2016effectiveness}. This powerful adversary (insider or outsider) can be a human or a trained bot. In imitation attacks, the attacker wears the victim's smartwatch either via after stealing it or the victim can leave it behind for a while and the attacker can try to impersonate the victim. On the other hand, the statistical attack is more complex and requires special tools and skills. Hence, WACA also considers these powerful attack scenarios in its adversary model. 
\end{itemize} 


The security evaluation of these attack scenarios and how WACA is robust against insiders, imitators, and statistical attackers are explained more in Sections V-A and ~\ref{sec:attacks}.

%% file: waca_architecture.tex
\section{WACA Architecture} \label{sec:arc}
In this section, we present the details of the WACA framework, which is a typing-based continuous authentication system using the accelerometer and gyroscope sensors of a smartwatch. 
Note that the WACA framework is complementary to the  first factor authentication mechanisms and it is flexible to work with any first factor,  
including one of the password-, token-, or biometric-based systems. Note that the first factor authentication is beyond the scope of this work. 

\subsection{Overview}
WACA consists of four main stages: \textit{Preprocessing, Feature Extraction, User Profiling, and Decision Module}. These stages, which are shown in Figure~\ref{overall}, work as follows: First, the raw sensor data is acquired from a smartwatch (1) through an app installed on the watch. Then, the raw data is transmitted to the computer through a secure wireless channel and the rest of the stages are performed on the computer except that Authentication Server (AS) is located in a trusted place. As the collected data includes a certain level of noise, in the preprocessing stage, the raw data is cleaned up by filtering (2) and transformed into a proper format for the next stages. The incoming data is used to extract a set of features (3). This set of features, namely \textit{feature vector}, represents the characteristics of the current user profile. In the enrollment phase (9), the created feature vector is stored in the AS. In the verification phase (4), the queried user profile is dispatched from the AS to the decision module (10, 11). The decision module computes a similarity score between the returned profile and the provided profile for the current user to make a binary authentication decision (match/no match). If the decision is a no match (5), then the user's access to computing terminal will be suspended and the user will be required to re-authenticate using the primary authentication method (e.g., password). However, when the decision is a match (6) then the user's access will be maintained. 
The profile of the current user in the AS will be updated after the correct match of the user profile (7). In WACA, this update frequency is a system parameter and can be set by the admin in the security policy. 
In this way, the user profile will be kept up-to-date over time. Whenever a typing activity is initiated on the keyboard of the computer, the smartwatch will be notified (8) again by the terminal to start over the authentication process continuously.
In the following subsections, we explain the details of WACA and its key stages.

\subsection{Data Collection}
In WACA, \textit{data collection} refers to capturing sensor readings from the user's smartwatch through a secure wireless communication channel (i.e., via WiFi or Bluetooth). An app is installed on smartwatch to listen to the physical sensors. Then, the raw sensor data is transmitted to the computer through a secure communication channel. 
\par Each row of the collected raw data of accelerometer is represented in the format of $\vec{acc}=<t_a,x_a,y_a,z_a>$ and gyroscope is represented as $\vec{gyro}=<t_g,x_g,y_g,z_g>$, where $t$ stands for timestamps and $x,y,z$ represent the different axis values of the accelerometer and gyroscope sensors. Each of $t,x,y$, and $z$ is stored as a different vector. 
The length of the vectors directly depends on sampling rate of the sensors and the time interval of the data collection. In WACA, the parameter \textit{sample size} refers to the length of these vectors and it is set as a configurable parameter while the parameter \textit{sample rate} is a constant system parameter that is characterized by the wearable device and app.

\subsection{Preprocessing}
In WACA, the \textit{preprocessing} stage refers to preparation of raw sensor readings for the next stages. It consists of cleaning and transformation of the raw data. In the cleaning part, the noise is removed. In order to remove the effect of the noise from data, we apply M-point Moving Average Filter (MAF)~\cite{wei1994time}, which is actually a simple low-pass filter and it operates by taking the average of M neighbour points and generates a single output. M-point filtering in equation form can be expressed as follows: $y[i]=\frac{1}{M} \sum_{j=0}^{M-1} x[i+j]$, where $x$ is the raw sensor data, $y$ is the new filtered data, and $i$ indicates the current sample that is averaged. The filtered data becomes smoother than the raw data without altering the value at that point. 

\par After filtering the noise, the data is transformed into appropriate forms for the next stage. Particularly, different types of sensor data are separated according to an assigned ID number during the sensor registration and then $x$, $y$, and $z$ axes of the sensor values are recorded as different vectors e.g., $\vec{x_a} = <{x_a}^1,...,{x_a}^n>$ and $\vec{x_g} = <{x_g}^1,...,{x_g}^n>$ for a profile of $n$ samples.


\subsection{Feature Extraction \& User Profiling}
In WACA, \textit{Feature Extraction} (FE) refers to the transformation of the time series raw data into a number of features. These extracted features will be used to create the feature vector of the user. The feature vector is the summary of the profile, which is checked in the decision module stage.

\par In order to create the feature vector, each feature is computed using the data vectors. As an example, the first feature is calculated from a function $f$, i.e., $f_1=f(x_a,y_a,z_a,x_g,y_g,z_g)$ and the second feature is calculated from another function $g$, i.e., $f_2=g(x_a,y_a,z_a,x_g,y_g,z_g)$ etc. Then, the final feature vector $\vec{f}=<f_1,f_2,...,f_n>$ is generated using all the calculated features.

\par As each element of the feature vector has different ranges, some of the features can be dominant in the distance measurement. To prevent this and create a scale-invariant feature vector, we apply a normalization to the feature vector to map the interval [$x_{min},x_{max}$] into the unit scale [0,1]. We formulate this linear normalization process in WACA as follows: $x_{new}=\frac{x-x_{min}}{x_{max}-x_{min}}$, $x_{min}$ and $x_{max}$ are the minimum and maximum value of the features of the user's enrolled templates.

\par After generating the final feature vector $\vec{f}$, in the user profiling stage, a user profile $\vec{p}$ is generated by adding the user ID and start and end timestamps of the data sample, i.e., $\vec{p}=<userID,t_{start},t_{end},\vec{f}>$. If the user is in the enrollment phase, this profile is transmitted to the AS to be stored in a database. Finally, if the user is unknown and a typing activity notification comes from the computer, the profile is passed to the Decision Module.


\begin{table}
\centering
\caption{Feature set extracted from sensor data in WACA. \label{feature}}
\resizebox{.48\textwidth}{!}{
    \begin{tabular}{|l|m{4cm}|c|}
    \hline
    \textbf{Domain}         & \textbf{Feature}                                                                                                                                                                        & \textbf{Length}  \\ \hline \hline
    Time           & Mean, Median, Variance, Average Absolute Difference of Peaks, Range, Mode, Covariance, Mean Absolute Deviation (MAD),  Inter-quartile Range (IQR), correlation between axes (xy, yz, xz), Skewness, Kurtosis & 12*6=72 \\ \hline
    Frequency      & Entropy, Spectral energy                                                                                                                                                        & 2*6=12    \\ \hline
    \textbf{Total \# of Features} & ~                                                                                                                                                                               & \textbf{84}      \\ \hline
    \end{tabular}
    }
\end{table}

The feature set used in our framework is presented in Table~\ref{feature}. These features were chosen as they performed well in similar contexts~\cite{gascon2014continuous,chernbumroong2011activity,killourhy2009comparing,serwadda2013verifiers}. Note that, though, WACA uses both time and frequency domain features. The feature set includes simple statistical metrics such as mean, median, variance of each 3-axis of the sensors readings and also advanced similarity metrics like covariance and correlation coefficients between the axes. While the statistical metrics are early indicators to show the tendency or intensity of the set, the similarity metrics help to differentiate between the users making hand motions in one or multiple directions when typing. Moreover, absolute of peaks is used to calculate average timing information between successive keystrokes. 
Covariance of two random variables $X$ and $Y$ in WACA is calculated as follows: 
$
covariance(X,Y)= E([X-E(X)][Y-E(Y)]),
$
where $X$ and $Y$ corresponds to the different axis of the accelerometer and gyroscope and $E()$ is the expected value of the variable in the WACA framework.
Similarly, the correlation can be calculated from the covariance as follows:
$
correlation(X,Y)=\frac{cov(X,Y)}{\sqrt{var(X)var(Y)}}.
$
Particularly, correlation and covariance of the pairs $(x_a,y_a)$, $(x_a,z_a)$, and $(y_a,z_a)$ are calculated for accelerometer, then, the same process is repeated to obtain the features of the gyroscope data.
\par In addition to these time domain features, the set also includes frequency domain features like entropy and spectral energy. The frequency domain features are based on the different typing frequency (period) behaviours of users. The different frequency impacts the energy calculation corresponding to spectral energy and randomness of the feature set corresponding to the entropy. In WACA, spectral energy of a signal is easily calculated from Fast-Fourier Transform (FFT) of the signal. FFT generates the frequency content of the time domain stationary signals over an interval. FFT of a signal is represented using complex numbers, which includes both amplitude and phase information. The spectral energy in WACA is calculated as follows: $energy(X)=\frac{\sum |FFT(X)|^2}{n}$, where $n$ is the number of FFT points and $||$ refers the magnitude of the given complex value of the sensed data. In WACA, the spectral energy is used to differentiate the users with different dominant frequency in both acceleration and torque while typing. Different dominant frequencies generate different spectral energies. On the other hand, the entropy is the measure of uncertainty and a higher entropy means more flatness in the histogram of the frequency distribution. The entropy of a random variable $X$ of values $x_1,x_2,...,x_n$ with the probabilities $p_1,p_2,...,p_n$ is calculated as follows: $entropy(X)=-\sum_{j=1}^{n} p_j \log_2 (p_j)$. In WACA, the entropy will discriminate between the users with random motion and the ones following similar pattern during the entirety of the typing activity.

\subsection{Decision Module}\label{subsec:decision_module}
The next stage in WACA is the \textit{decision module}. The task of this stage is classifying the user as authorized or unauthorized for given credentials entered during the initial login. For the purpose of authentication, we use distance measures. 
The distance measure methods simply calculate the distance between two vectors or data points in a coordinate plane. It is directly related to the similarity of compared time-series data sets. The most widely used distance measure is \textit{Euclidean Distance}. It is actually just the distance between two points in vector space and is the particular case of \textit{Minkowski Distance}, which is expressed as follows:
\begin{equation}
    distance(\vec{x},\vec{y})=(\sum_{i=1}^{n} (x_i - y_i)^p)^{\frac{1}{p}},
\end{equation}
where $\vec{x}=(x_1,x_2,...,x_n)$  and $\vec{y}=(y_1,y_2,...,y_n)$ are the set of sensor observations to be compared. If $p=2$, it is Euclidean distance and has been extensively used in the keystroke-based authentication methods~\cite{alsultan2013keystroke}. WACA calculates the distance and returns the result by comparing it with a configurable predetermined threshold value (i.e., genuine if $distance<threshold$, impostor if $distance\geq threshold$).

In addition to Euclidean and Minkowski Distances, there are several distance measurement methods utilized in biometric authentication systems which may perform differently depending on the context. Therefore, we also tested different distance metrics in our experiments to see which performs the best for WACA. Other distance metrics that we tested in our experiments are \textit{Cosine Distance, Correlation Distance, Manhattan (Cityblock) Distance} and \textit{Minkowski with p=5}. The performance of each one is given in Section~\ref{subsec:auth}.

%% file: evaluation.tex
\section{Performance Evaluation} \label{sec:eval}
In this section, we evaluate the efficacy of WACA using data from real users. We also explain the details of our experiment setup, methodology, and performance metrics, and discuss the results. 
While evaluating WACA, we considered three evaluation metrics, which are as follows:
\begin{enumerate}
    \item \textit{How accurately can it differentiate between genuine and impostor users?}
    \item \textit{How fast can it  detect an impostor?}
    \item \textit{How accurately can it identify an insider?}
\end{enumerate}
Specifically, we primarily focus on measuring the detection accuracy and speed of WACA. We, first, conduct authentication experiments. In these, we measure how WACA performs when users type a different or the same text. We also analyze how the sample size and the detection technique impact WACA's performance. Second, we measure how successful WACA is against insider threats. Third, we test the robustness of the framework against more advanced attacks, imitation and statistical attacks. Finally, we evaluate the overhead of WACA. 


\noindent \textbf{Data and Collection Methodology.} 
%
In our experiments, we collected data from 34\footnote{\revision{Not all of them participated in all experiments.}} human subjects whose ages ranged from 18 to 38  with a mean of 25.5 and standard deviation of 4.1 using two different smartwatches. 
The detailed information about participants are given in Table~\ref{demo}. Furthermore, note that our research study with the human subjects was conducted with the appropriate Institutional Review Board (IRB) approvals.

\begin{table}[htbp]
\centering
    \caption{Characteristics of participants \label{demo}}
\resizebox{.8\columnwidth}{!}{%
    \begin{tabular}{lll}
    \toprule
   \multirow{2}{*}{Age}                   & Mean         & 25.4 \\
    ~                     & Std          & 4.1  \\  \hdashline[1pt/10pt]
     \multirow{2}{*}{Gender}                 & Male         & 22   \\
    ~                     & Female       & 12   \\ \hdashline[1pt/10pt]
    \multirow{2}{*}{Smartwatch Brand}      & Samsung Gear Live      & 20   \\
    ~                     & LG G Watch R          & 14   \\ \hdashline[1pt/10pt]
    \multirow{2}{*}{Watch Hand Preference}  & Right        & 7    \\
    ~                     & Left         & 27   \\ \bottomrule
    \end{tabular} }

\end{table}   


 \begin{figure*}[t!]
    \centering   
\begin{minipage}{0.48\textwidth}
        \centering
        \includegraphics[width=1\textwidth, trim=0cm 0cm 0cm 0cm]{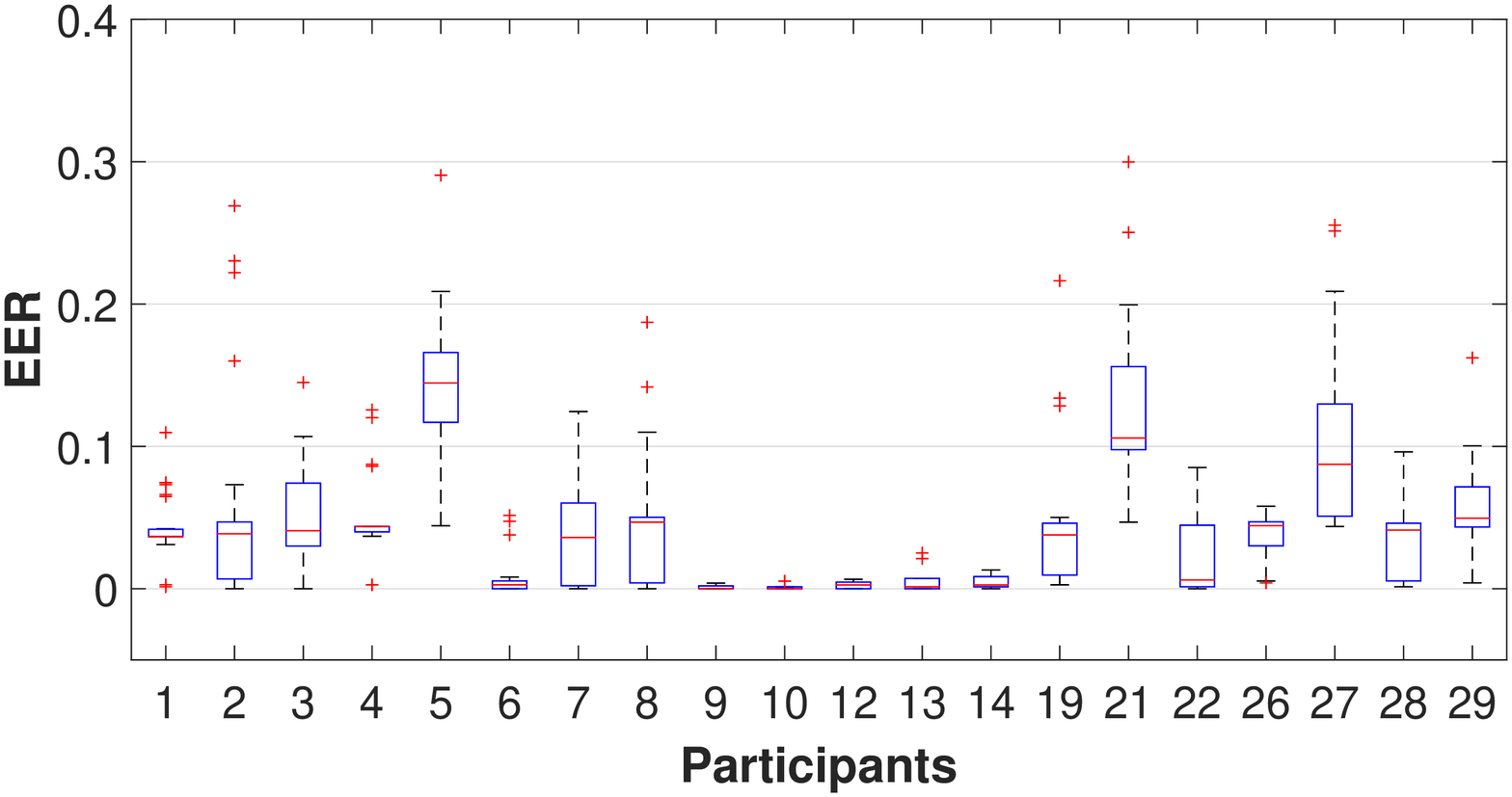} 
         \caption{ EER for each participant with a sample size of 1000 using Manhattan (Cityblock) distance metric during Typing Task-1. Average EER is  0.0513.\label{typing_err_free_text2}}
     \end{minipage}\hfill
    \begin{minipage}{0.48\textwidth}
        \centering
        \includegraphics[width=1\textwidth]{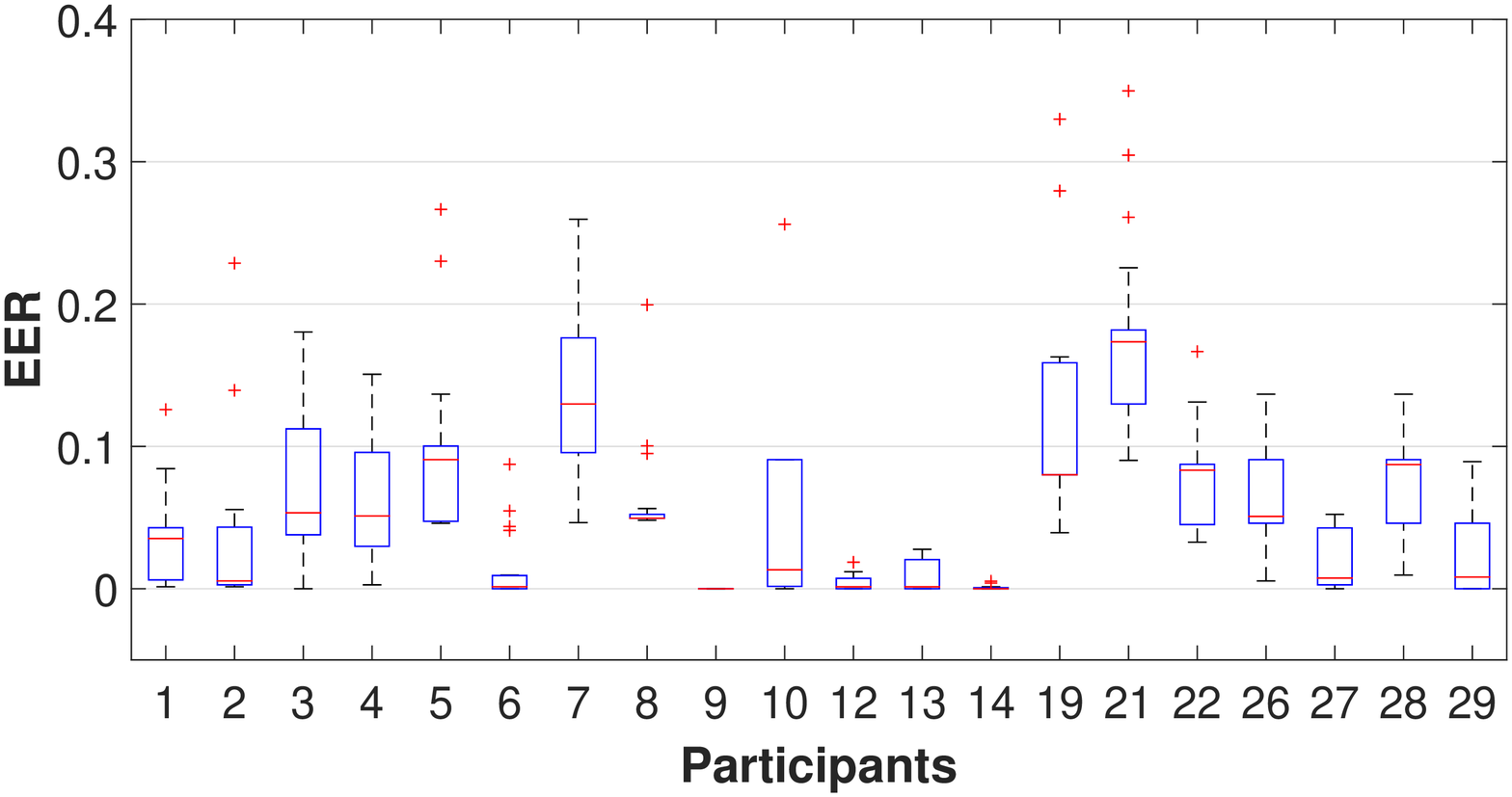} 
       \caption{ EER for each participant with a sample size=1000 using Manhattan (Cityblock) distance  metric during Typing Task-2. Average EER is  0.0647.~\label{eer_cosine}}
       \end{minipage}    
\end{figure*}

During the collection of data, an Android Wear smartwatch with an installed data collection app was distributed to the participants and the participants were asked to type a 
text 
while the program in the smartwatch was recording its sensory data. 
The participants were free to choose the hand (left/right) on which they wore the smartwatch.   
Moreover, they were also given the freedom to adjust the sitting position and the keyboard and screen position according to their comfort levels. Throughout these experiments, we utilized 
a standalone qwerty keyboard to have generic results. Before typing each text, the participants were also given enough time to read the texts to make them familiar with the text as typing a familiar text is a more common activity. 


\par The participants were involved in three typing tasks conducted in three different sessions. They were asked to type with their normal typing style without noticing that their data was recorded. The three data sets were compiled as follows:

\begin{itemize}
    \item \emph{Dataset-1, Typing Task-1:} The participants were asked to type a story from a set of short and simple stories from the American Literature\footnote{https://americanliterature.com/100-great-short-stories} for four minutes. The story was chosen randomly by the participants. On average, four minutes of data corresponds to
    25000 readings for each participant (Total: 850000 readings). 
    \item \emph{Dataset-2, Typing Task-2:} For this data set, all the  participants were asked to type the same text\footnote{https://en.wikipedia.org/wiki/The\_Adventures\_of\_Tom\_Sawyer} for four minutes. For each participant, almost the same amount of data is collected as Dataset-1. This dataset is important to be able to measure the quality of the features. 
    
    
    \item \emph{Dataset-3, Typing Task-3:} The participants were instructed to imitate someone else' (victim) typing pattern by watching the prerecorded video of the other person. For these experiments, one of the participants was recorded on video while typing 
    from a perspective that the hand motions, smartwatch, keyboard, and the screen could be seen. Although it was not required, the perspective allowed to infer what the victim was typing by watching. This dataset was primarily used to analyze the attacking scenarios.  
\end{itemize}

Note that in all the experiments, the datasets obtained from all these tasks 
were always used by 
cross-validation techniques (i.e., partitioning the data set into randomly chosen two sets for training and testing). 
Therefore, even if the same text was typed by all the participants in Typing Task-2, the compared samples always corresponded to different texts for a participant.


In order to characterize the participants' typing performance, 
the typed text was also recorded, in addition to the sensor data, and was analyzed. In this regard, the participants' typing speed was also measured. We observed the typing speeds between 20 and 75 words per minute (wpm) on average.

In our experiments, we split the collected data sets into equal size chunks, called \textit{sample size}. It is the number of samples (i.e., row) in a chunk. Each chunk consists of 8 columns of data, two of which are timestamps and the others are 6 dimensional sensor data. Sample size is the main system design parameter in our experiments as it has a direct impact on the time required to collect data. Particularly, the time $t$ required to collect data with the sample size can be represented as $t=sample \ size/100$ in seconds as the sampling rate in our experiments was $100Hz$.
    
As discussed in Section~\ref{subsec:decision_module}, the decision module in WACA computes similarity of an unknown sample with the one stored in the authentication server. 
Then, in order to decide whether an unknown user is accepted or rejected, the decision module compares a similarity score, a value of $[0,1]$, with a predefined parameter threshold. In our experiments, we also tested WACA's decision module with five different distance measuring techniques and reported the performance of each technique.  

 \noindent \textbf{Performance Metrics.} In the authentication experiments, we used \emph{Equal Error Rate} (EER) as it is a commonly accepted metric to assess the accuracy of WACA. EER is calculated using two metrics: False Acceptance Rate (FAR) and False Reject Rate (FRR). FAR is the rate of incorrectly accepted unauthorized users among all the illegal attempts: The increase in FAR is a direct threat to system's security level. For more valuable assets, increasing the threshold will decrease FAR. On the other hand, FRR is the rate of incorrectly rejected authorized users among all the legitimate authentication attempts. Contrary to FAR, FRR can be decreased by decreasing the value of threshold. Finally, EER is the point that gives the closest FAR and FRR point for a given threshold (ideal EER is the intersection point of FAR and FRR) and the lower the EER the better is an authentication system.

\subsection{Results}  \label{subsec:auth}
In this section, we present and discuss the evaluation results.

\noindent \textbf{Impact of the text dependency.} 
In this experiment, our goal is to analyze how EER changes among the participants. We try to answer the question: \textit{How does WACA perform with the typed text?} This is also a more advanced analysis of the framework and the fundamental idea than that of in Section 2. 

Specifically, for this experiment, we used Typing Tasks 1 (any text) and Typing Task 2 (the same text) datasets. For each sample of a particular user, we computed the differences from other users' samples. For this purpose, we computed the $NxN$ dissimilarity matrix, where $N$ is the total number of samples for all the participants. The dissimilarity matrix was calculated by measuring the similarity of each sample to all the other samples using leave-one-out cross-validation method~\cite{friedman2001elements}. 

Then, for a given threshold and participant, the ratio of the rejected and accepted samples were computed to obtain FRR and FAR, respectively.   
This process was repeated by incrementing the threshold by 0.01 in each step for all the samples of all the  participants. 
This gave us a set of EER for each participant. Note that in a real system, FAR/FRR rate can be tuned according to the system preferences, but here our purpose is to find 
an acceptable performance metric for WACA. The results are plotted in  Figure~\ref{typing_err_free_text2} for Typing Task-1 and Figure~\ref{eer_cosine} for Typing Task-2. 
%
%
Average EER for the Typing Task-1 experiment was 0.0513. 
Figure~\ref{eer_cosine} compares the EER of participants for the Typing Task-2 experiment. Average EER for this experiment was 0.0647. Another observation from the plots is that 
some participants have more distinctive typing characteristics than others using both the datasets. 
%
%

\begin{figure}[t]
    \centering
          \includegraphics[width=0.65\textwidth, trim=10cm 0cm 0cm 0cm]{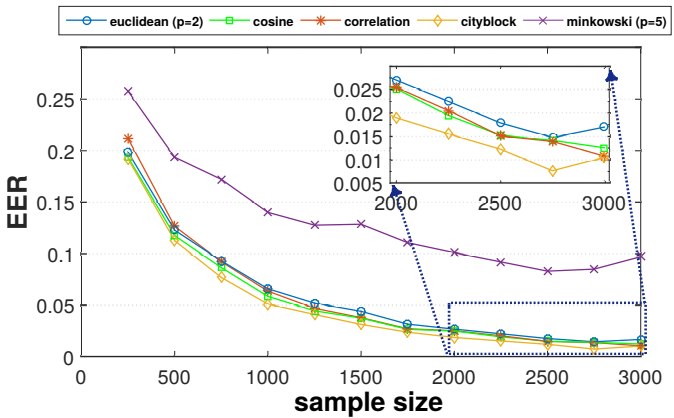}
          \caption{Average EER according to different sample sizes using different distance metrics while users are performing Typing Task-1. \label{compare_free}}
\end{figure}

If we compare the ERR of each participant in both the experiments, we see that they are also close to each other, where a few of the participants perform very distinctive behaviours (e.g., participant 21). However, the overall distribution of EER over the participants is similar in both the experiments. Recall that in Typing Task-1, all the participants typed different texts, while they typed the same text in Typing Task-2.

\textit{Overall, in this analysis we report the average EERs of both  the experiments are close (around \%1), which supports the usability of WACA regardless of the typed text for the continuous authentication session.}

\begin{figure}[t]
    \centering
          \includegraphics[width=0.65\textwidth, trim=7cm 0cm 3cm 0cm]{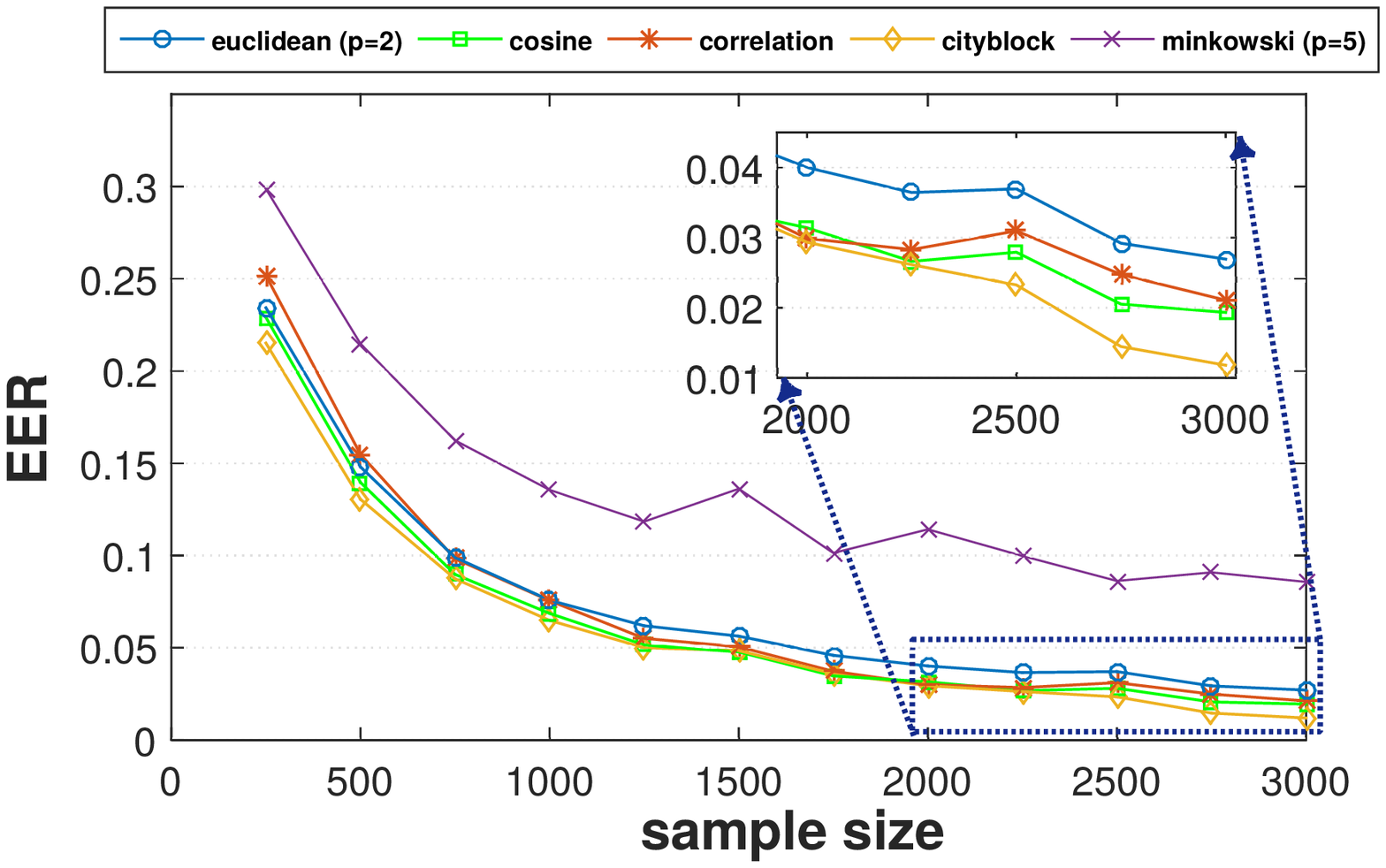}
          \caption{Average EER according to different sample sizes using different distance metrics while users are performing Typing Task-2. \label{compare_same}}
\end{figure}

\noindent \textbf{Impact of the sample size and the distance measuring technique.} 
In these experiments, \textit{our goal was to assess how different sample sizes and the distance measuring techniques used in WACA impact the performance.} 
For this, we varied the sample size from 300 to 3000 and utilized five different distance measuring techniques, Euclidean (p=2), Cosine, Correlation, Cityblock, and Minkowski (p=5).  
Again, two types of participant datasets, Typing Task-1 (any text) and  Typing Task-2 (the same text), were used.   
Figure~\ref{compare_free} (Typing Task-1) and Figure~\ref{compare_same} (Typing Task-2) present the main results when the sample size increases.

As can be seen in Figure~\ref{compare_free} when the participants typed different texts, the EERs are generally decreasing with the increase of sample sizes as expected.  The EERs go under 0.05 after the sample size of 1500 for all the distance metrics utilized except for  Minkowski (p=5). Then, the EER is converging to the value of 0.01-0.02 through the sample size of 3000. In the best case, EER 0.007 is achieved with the the sample size of 2750 for the Manhattan (cityblock) distance measurement technique. 

Figure~\ref{compare_same} presents the results of the same-text experiment (Typing Task-2). As in Figure~\ref{compare_free}, the general behavior is that the EERs are decreasing with the increase of the samples. The lowest EER of 0.01 is achieved using the Cityblock distance measuring technique at 3000. We also see the convergence of EER in Figure~\ref{compare_same} as  Figure~\ref{compare_free}. Plots are starting to converge around sample sizes 1500-2000 and converging to 0.01 for Cityblock and Correlation distance measuring techniques. We also see that at 3000, 0.02 EER is obtained for Cosine and Correlation techniques. However, if shorter data collection time is of interest, a sample size of 2000, which needs 20 seconds for data collection, gives 0.03-0.04 EER. However, if we increase the sample size, both the accuracy and the data collection time are increasing. This means the time needed to catch an adversary or more generally the re-verification period would also increase. Therefore, an optimal sample size should be adjusted according to the preferences in a  real application based on the usage needs or the security policies. 

\textit{To conclude, 
the features in WACA can successfully differentiate the users from their typing rhythm with a very small error rate (1\%) independent of the typed text. 
There is a natural trade-off between the EER and data collection time, which should be configured according to the security needs of an organization.} 


\noindent \textbf{The accuracy of insider threat identification.} As noted earlier, the insider threat detection is important in continuous authentication systems as a potential attacker is likely to be an insider. In order to effectively locate such an insider attacker within an organization where WACA is employed, an identification mechanism is needed. Hence, WACA includes \textit{Multilayer Perceptron algorithm} (MLP)~\cite{wasserman1988neural} to defend against and identify insider threats. MLP is a feedforward neural network model which maps a set of input data into a set of outputs through the interconnected processing elements (neurones)~\cite{gardner1998artificial}. We used MLP for the task of identification. Identification is a one-to-many classification task and requires a training set. We assume that the insider's data is also stored in the authentication server's database (training set) as a legitimate user.
We used MLP since it gave the best results in our experiments.

In order to analyze the efficacy of WACA against insider threats, we analyzed the impact of the sample size and the size of the training data on accuracy. For this, we focused on two test scenarios that could be relevant in real investigations:  \textit{Scenario 1:} In the first scenario, we built our test model using the same text and tested again using the same text with the $5$-fold cross validation technique. For this scenario, we utilized Typing Task-2 Dataset for both the training and testing. This type of scenario can be useful as all the users are asked to type a provided text and during the investigation, all users are asked again to type the same text. The results are presented in Table~\ref{iden_same}. \textit{Scenario 2:} In the second scenario, the test model was trained with the same text dataset, which is the same for all the participants and tested using random-text experiments, where each user typed a randomly chosen text. For this scenario, we utilized Typing Task-2, Typing Task-1 Datasets for training and testing, respectively. This scenario is suitable for cases where all the users are enrolled using the same text, but a user is verified while typing a random text. The results for this test scenario are presented in Table~\ref{iden_diff}.  

As can be seen in Table~\ref{iden_same}, in the best case, 99.2\% identification rate of an insider threat can be achieved with the sample size of 1500 while the model is trained with 5 samples.  Even with 2 samples of the insider, 93.7\% accuracy rate can be achieved with the sample size of 1500. 

\begin{table}[!ht]
    \centering
        \caption{Scenerio 1: The same text is used for both training and testing using the Multilayer Perceptron algorithm. \vspace{-0.25cm}\label{iden_same}}
        
     \vspace{.2cm}
     
       \resizebox{.9\columnwidth}{!}{%
    \begin{tabular}{cc}
    \begin{tabular}{cccccc}
    \multicolumn{6}{c}{\centering \textbf{The accuracy of insider threat identification ($\%$)}} \\ \hline \hline
       \multicolumn{6}{c}{\centering Training Set}         \\  \cline{3-4} \bigstrut
      Sample size & 1     & 2     & 3     & 4     & 5  \\ \hline \bigstrut
     
    1500 & 77.8  & \textbf{93.7}  & \textbf{97.2}  & \textbf{98.4}  & \textbf{99.2} \\   \bigstrut
   1000 & 62.8  & 87.6  & \textbf{93.8}  & \textbf{95.3}  & \textbf{97.1}  \\  \bigstrut
    500 & 37.5  & 63.7  & 75.9  & 83.1  & 89.6  \\  \bigstrut
    250 & 28.5  & 43    & 53.1  & 61.8  & 62.1  \\  \bottomrule
 
    \end{tabular}
    \end{tabular}
    } 
\end{table}

Scenario 2 aims to answer the question of "\textit{Can an insider be identified while typing a random text even if s/he is enrolled while typing a given text ?}" Table~\ref{iden_diff} presents \noindent the result of this question for Scenario 2. As can be seen from Table~\ref{iden_diff}, similar to Scenario 1,  the accuracy rates increase as the sample sizes and training set increase, and the time to build model and time required to catch the attacker is also increasing. 3 training samples and the sample size is 1500 or 4 training samples with the sample size of 1000 may be the two most optimal choices for real cases.
\begin{table}[!th] 
    \centering
    \caption{Scenerio 2: All users are trained with the same text and tested with random texts using the Multilayer Perceptron algorithm. \label{iden_diff}} \vspace{-0.25cm}
      \resizebox{.9\columnwidth}{!}{%
    \begin{tabular}{cc}
    \begin{tabular}{cccccc}
     \multicolumn{6}{c}{\centering \textbf{The accuracy of insider threat identification ($\%$)}} \\ \hline \hline \bigstrut
     & \multicolumn{4}{c}{\centering Training Set} &  \\  \cline{3-4} \bigstrut
     Sample size   & 1     & 2     & 3     & 4     & 5     \\ \hline  \bigstrut
   
    1500  & 55.8  & 80.1  & 88.7  & 89.8  & \textbf{91.8}  \\ \bigstrut
    1000  & 51.7  & 82.7  & 83.2  & 86.1  & 86.8  \\ \bigstrut
    500   & 29.9  & 51.3  & 66.7  & 73.8  & 76.5  \\ \bigstrut
    250   & 22.1  & 33.6  & 41.9  & 49.8  & 54.1 \\ \bottomrule \bigstrut
        \end{tabular}
    \end{tabular}}
    \vspace{-.5cm}
 \end{table}

\emph{Overall, WACA can achieve 0.01 error rate with almost 30 seconds of the data collection (see Figure~\ref{compare_free} and~\ref{compare_same}) in the best case. If a shorter time is of interest, 0.02 error rate is achieved with 20 seconds of the data collection.  Moreover, if 5 training samples with 1500 sample sizes are obtained from a potential insider threat, WACA could identify the insider with 99.2\% accuracy rate while typing the provided text (see Table~\ref{iden_same}) or with 91.8\% accuracy rate while typing a random text (see Table~\ref{iden_diff}).}

%% file: advanced_attacks.tex
\subsection{Advanced Attacks on WACA with More Powerful Adversaries} \label{sec:attacks}
In this subsection, we evaluate the performance of WACA against two powerful attacks: imitation~\cite{tey2013can,huhta2015pitfalls} and statistical~\cite{serwadda2013examining,stanciu2016effectiveness} attacks. In these attacks, the attacker is aware of WACA and can try to defeat WACA using special tools and skills. For this purpose, we also developed generic attacking scenarios that can also be utilized by other future continuous authentication studies.

Various attacks against classical keystroke dynamics that exist in the literature can also be used to attack WACA. The attacker can be a human or a trained bot. A human-type attacker can perform \textit{zero-effort attacks\footnote{Also called \textit{zero-information attack}.}}~\cite{rattani2013biometric} or \textit{imitation attacks}~\cite{tey2013can} to defeat the WACA's authentication system. In \textit{zero-effort attacks}, the attacker tries to defeat the authentication system without any effort or prior knowledge. Zero-effort attacks will not be successful due to the low EER values in WACA as analyzed in the previous sub-sections. However, the effectiveness of the imitation attacks performed by a human should be investigated as noted in some recent studies~\cite{tey2013can,huhta2015pitfalls}. 

In addition to these attacks, another recent attack against the behavioral biometrics~\cite{serwadda2013examining,stanciu2016effectiveness} has emerged, which is called \textit{statistical attacks}. In this attack, a bot is first trained using typical user data from a large population. Then, the bot generates random permutations of the features to mimic a legitimate user. In addition to human and robot attacks, a \textit{replay attack} using a key-logger~\cite{giuffrida2012memoirs} is noted in the literature, which can also be performed against the keystroke dynamics. However,  a key-logger installed on the computer can obtain only the timing of the keystrokes, which is solely not enough to use it in a replay attack against WACA as there is not a way that a key-logger can obtain the three dimensional sensor data collected by the smartwatch. 
In the next sub-sections we consider these two powerful attacks (imitation and statistical) and investigate the effectiveness of WACA against them. In these cases, the attacker would have somehow obtained the victim's smartwatch or manipulates his smartwatch. We use the zero-effort attacks as a baseline to evaluate the success of the imitation and statistical attacks. In imitation attacks, the attacker either steals the victim's smartwatch or the victim can leave it behind for a while, then the attacker wears the victim's smartwatch and can try to impersonate him while attacking. On the other hand, the statistical attack is more complex and requires special tools and skills. In this type of attack, we assume the attacker can provide its input data to the system. It manipulates its username and profile data to get an access to the computer that he does not have permission.

\subsubsection{Imitation Attacks} \label{subsec:attack1}
In this subsection, we evaluate the performance of an imitating adversary, who knows that WACA is already installed on the current system. The adversary is assumed to be watching his victim by standing nearby or trying to imitate the victim's typing style by looking at the previously recorded video of the victim while typing. S/he is also assumed to be opportunistically waiting for the right time to mimic the victim.
\begin{figure} 
    \centering
          \includegraphics[width=0.4\textwidth,trim= 0cm 0cm 0cm 0cm]{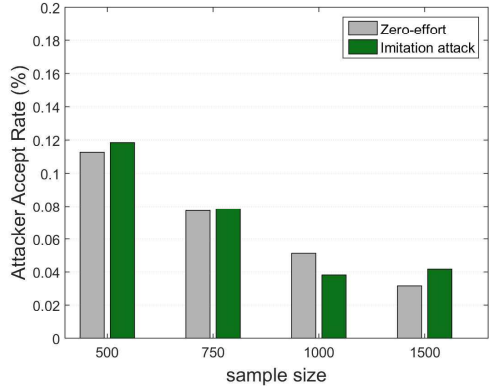}
          \caption{ Attacker accept rates for different sample sizes. The results show that even though an imitation attacker is aware of WACA in a targeted system, s/he has no more advantage than a zero-effort attacker.  \label{imitation}}
\end{figure}
In order to replicate this imitation attack scenario, we recorded a 15 seconds video of a legitimate user and presented this video to an attacker (i.e., another participant in our experiments). The video showed the user as s/he was typing and thus the hand, fingers, watch and keyboard were all clearly visible. By watching the video (multiple times allowed in experiments), the attacker tried to imitate the legitimate user. Note that this scenario would increase the chances of a successful attack when compared to a real-life case where the attacker would possibly only have limited opportunity to watch a victim. We also collected the victim's typing data to evaluate the performance of the attackers. We computed EER for this attack scenario and compared it with the case when there was a zero-effort for the attack. 
%
In the zero-effort attack, we used the data set obtained in Typing Task-1 Dataset. We applied the leave-one-out method~\cite{friedman2001elements} by leaving the victim's data out as in the other authentication experiments 
while calculating EER (i.e., the intersection of FAR and FRR) of the victim. In the imitation attack, since we only had the impostor attempts, EER would be equal to the attacker's acceptance rate. We also note that WACA was directly tested without any change. The results are presented in Figure~\ref{imitation}.

As presented in Figure~\ref{imitation}, the attackers have different success rates (attacker accept rate) for different sample sizes. The highest success rate was achieved when the sample size is equal to 500, but the success rates are decreasing to much lower rates as the sample sizes increase. A sample size of 500 corresponds to almost 2-3 keystrokes for the sampling rate used, which is not enough to measure and settle down for some of the features. So, this is not practical from the attacker's perspective. Beyond 1500, which corresponds to 15 seconds of sensor readings, the probability of an attacker to imitate a user is significantly decreasing (i.e., 0.04). \textit{These results indicate that even though an attacker is aware of WACA in a targeted system,  s/he still has a very low chance to be successful.}  

\begin{figure}[!t] 
    \centering
          \includegraphics[width=0.4\textwidth,trim= 0cm 0cm 0cm 0cm]{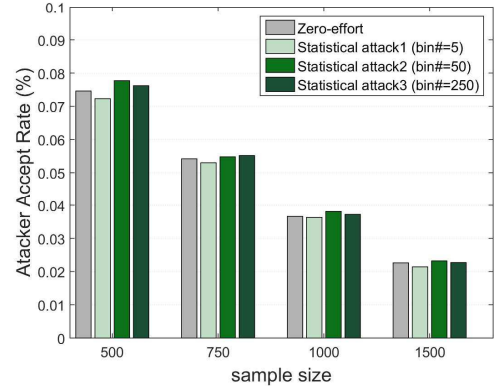}
          \caption{ 3 different statistical attacks against WACA with different sample sizes. These results show that despite the small increase compared to zero-effort, the attacker does not have a chance to defeat WACA using the systematically generated fake data. \label{stat_attacks}}
          \vspace{5pt}
          \end{figure}
\subsubsection{Statistical Attacks} \label{subsec:attack2}
In this subsection, we evaluate WACA against statistical attacks. In this attack scenario, it is assumed that the attacker has a database obtained from AS consisting of the user profiles. Similar to the imitation attack, it is also assumed that the attacker has a capability to provide its input to system. As mentioned earlier, this can occur either by obtaining the victim's smartwatch or if the attacker is an insider, it can manipulate its own input data to deceive WACA. It is also worth mentioning that we assume the attacker has only a limited amount of time to attack; therefore, it only tries the data that has the highest chance to get in, which we will refer to as $topBins$ in the attack algorithm that will utilized and noted below.

Note that statistical attacks are very powerful attacks and it is successfully implemented to bypass the conventional keystroke-based systems~\cite{serwadda2013examining}. It is based on the generation of fake (synthetic) inputs using common features of a given population. The idea behind this attack is using the random combination of the most frequent features of the population to defeat the authentication system. We designed the following attack scenario to test WACA against the statistical attacks. 
%

In our attack, we used both Typing Task-1 and Typing Task-2 datasets as input. Each participant was chosen as a victim iteratively and the other participants' samples were used to generate forged data samples. Then, the forged samples were used to attack the victim. For this, a histogram was generated for each feature of all the participants in the datasets except the victim. The forged samples were generated as in Algorithm~\ref{alg:stat} in Appendix~\ref{sec:stat}. Overall, we created three different statistical attackers with three different capabilities (bin sizes in the histogram). Statistical Attacker-1, Statistical Attacker-2, and Statistical Attacker-3.  
%
%
Before running the algorithm for attacking on WACA, we first calculated the EER for each user without adding any forged data. Similar to the imitation attacks, the attacker acceptance rate in zero-effort attack corresponds to the average EER. We conducted experiments without attack under varying sample and bin sizes. The results are shown in Figure~\ref{stat_attacks}. 

In Figure~\ref{stat_attacks}, we can see that bin number 50 has the most successful result on attacking victims. This is because if we increase the bin number and keep the bins with the highest number of occurrences constant, the width of the bins will narrow; so, the range of the forged data will be confined to a very small range. On the other hand, if we decrease the bin number significantly, the less frequently occurred bins will also be included in the sample generation range, which will decrease the success rate of the attacks. Finally, we note that in the
attack scenario, we choose each user in our dataset as a victim in an iterative way. \textit{These results show that despite the small increase compared to zero-effort, the attacker does not have a chance to defeat WACA using the systematically generated fake data due to its high dimensional feature vector in WACA's design.} 


\textit{As a summary, 
neither the imitation nor statistical attacks puts WACA in danger as their success rates are very close to zero-effort attacks. 
The strength of WACA is  from its high-dimensional feature vector and data originating from independent sensory sources.}

\subsection{Resource Consumption} \label{sec:res}
In WACA, a smartwatch, a computer, and an authentication server work together. In this subsection, we only analyze the resource consumption of relatively constrained smartwatches. It is worth noting that we monitored the consumption of our application while it was running continuously; however, in WACA the data collection app does not have to be running continuously. It can happen periodically or on-demand because the data collection runs only when the smartwatch is notified by the computer that the user is typing on. We analyse the performance of both LG G Watch R and Samsung Gear Live smartwatches used in the experiments. Both smartwatches have Cortex-A7 at 1.2GHz and 512MB RAM, but Samsung uses 300mAh battery, while LG is using 410mAh battery. The results are presented in Table~\ref{power}.

\begin{table}
\centering

\caption{Resource consumption of the smartwatches used in the experiments: \textit{LG Watch R} and \textit{Samsung Gear Live}.  \label{power}}
\vspace{-.1cm}
\resizebox{.45\textwidth}{!}{
    \begin{tabular}{lccc}
   
         &   & \textbf{LG G } & \textbf{Samsung}  \\  
          &   & \textbf{Watch R} & \textbf{Gear Live}  \\ \hline \hline  \bigstrut
    CPU (no WACA)         &    &  4.5\% &  4.5\%\\ \hline \bigstrut
    CPU (w/WACA)       &    &  7.5\% &  16.8\%\\ \hline \bigstrut
    Memory (no WACA)     &    & 4.5 MB &   4.5 MB \\ \hline \bigstrut
    Memory (w/WACA)     &    & 15.2 MB &   13.8 MB \\ \hline \bigstrut
    Battery     &10Hz  & 1.1\% & 1.2\% \\ \bigstrut
                     &30Hz  & 1.6\% &  0.3\% \\  \bigstrut
                     &100Hz & 2.1\% &  2.4\% \\ \hline \bigstrut
     Data Size  &~10Hz  & 0.3 MB &  0.3 MB   \\  \bigstrut
                     &30Hz  & 0.6 MB &  0.9 MB \\   \bigstrut
                     &100Hz & 4.1 MB &  6.5 MB \\ \bottomrule                 
                    \end{tabular}
    }
\end{table}

In all the experiments, 
we both monitored the memory and CPU resource utilization of the smartwatches in the default mode (i.e., not actively running any app - no WACA) and while the app was running (w/WACA). In the default mode, both smartwatches used almost 4.5MB memory and 4.5\% CPU  their consumption while the app was running as shown in Table~\ref{power}. As compared to the default memory usage (no WACA), the memory consumption in the smartwatch in WACA is increasing, but it is still at an acceptable rate.      


In addition to memory and CPU consumption, we also analyzed the power consumption and data size while running our app for 10 minutes. We excluded the power consumption of the screen as the screen can be turned-off or the smartwatch can be in the ambient mode during the data collection of WACA. We see that the power consumption of the app scales by the sampling frequency. However, when we decrease the sampling rate, the time needed to collect a certain amount of data will also increase. Hence, the optimum sampling rate should be tuned according to the desired security policy. 

%% file: discussion.tex
\section{Discussion} \label{sec:disc}

\noindent \textbf{\revision{Security Policy Implementation Considerations.}} WACA works by checking if the current user's profile matched the profile of the logged-in user. When an unauthorized access attempt is detected, the reaction depends on the  previously decided security policy. Depending on the \revision{security} policy, when an attacker is detected, the screen can be locked and the user can be challenged to re-login; the management and security teams can be alerted in real-time; or a notification email can be sent to the registered e-mail of the logged-in user, and so on. Moreover, we showed that WACA can  differentiate an insider from an outsider accurately. In suspicious cases, the administrator can do further investigation to detect the insider, and as we noted earlier, the insider detection is possible in WACA. We also note that even if WACA catches an insider attacker, WACA can not know if the attacker has the full key, which is out of scope this work. \revision{Therefore,} even if the system is logged-out, an insider can log-in again if it has the full key. Therefore, resetting the initial authentication factor should be considered in the security policy in this case. Finally, the server can also log the failed attempts to prevent from attacks aiming to drain the smartwatch's battery. 


\noindent \textbf{Privacy.}
In WACA, the computer and the wearable are the devices that belong to the user or belong to the same authentication realm and thus are trusted. The only device that may threaten the privacy is the AS. In WACA, after collecting the raw sensor data from the smartwatch, either the raw sensor data can be transmitted to the AS, or the features can be computed on the smartwatch and the the feature vector can be transmitted. If the raw sensor data is sent to the AS, the AS may try to infer the user's typed characters from the raw sensor data. The more secure way would be to compute the features on the smartwatch and to keep the feature vectors of the profiles of the users in the AS. In that case, 
the transmitted feature vector has only the mean of the values of the  multi-dimensional sensor data and thus inferring the typed characters would not be possible at the AS.

%% file: limitations.tex
\section{Limitations} \label{sec:limit}

\noindent \textbf{Absence of the biometric.} WACA needs to capture sufficient amount of sensor data when a user is typing in order to authenticate him/her as a legitimate user. \revision{Therefore, WACA can perform better for applications that heavily rely on users typing text such as word processors.} However, the user may not always be typing or the smartwatch may not always provide sensor data. Examples of such cases can occur when the user is watching a movie or scrolling through a document and thus not touching the keyboard. \revision{Therefore,} the biometric may not be always present. Indeed, the absence of the biometric is an issue for all the currently proposed continuous authentication methods~\cite{niinuma2010continuous}. In these cases, the straightforward solution is that the user can disable WACA for that duration. 
Disabling WACA will default to the initial security level of the system. Fortunately, WACA can provide a better solution for these inactive cases. If the smartwatch deployed for data collection has a proximity sensor (or using the Bluetooth signal strengths~\cite{huhta2015pitfalls}), the user's closeness to the computer can be checked. If the user moves away, then the user can be logged out. 
However, this approach would not be effective if the watch is stolen and worn by somebody else. 

\revision{Other than long time inactivity of the user, some individuals may read a book, use the mouse only, and enter short text snippets like emails. In these cases, the user will be typing rarely or typing only short texts. WACA can still increase the security of this kind of usage scenarios by combining it with the traditional time-out methods. The traditional time-out methods postpone the lock whenever an activity has been detected. This postpone mechanism can be replaced with WACA so that whenever WACA identifies the user, the locking can be postponed. In this way, the security of the system will not be relying on the blindly detection of activity; instead it relies on the identification of the user.}

\revision{Finally, the absence of the biometric will occur if the smartwatch' battery depletes and} the smartwatch batteries do not last long before one needs to charge them and note that most of the batteries have an estimated life of 1-3 days~\cite{best}. We provide a detailed resource consumption analysis in Section  ~\ref{sec:res}. 

\noindent \textbf{Abnormal cases.} 
\revision{In WACA, if the smartwatch data of the attacker is available and an interaction with the keyboard occurs, WACA detects the attacker with high probability. However,} there are two more cases which should be considered: 
\begin{enumerate}
    \item \textit{No interaction, but data:} Smartwatch provides data while there is no interaction with the keyboard.
    \item \textit{Interaction, but no data:} No smartwatch data is available while there exists an interaction with the keyboard. 
\end{enumerate}

In the first case, the currently logged-in user is nearby and doing something else, but not typing or typing in a nearby device not on the logged-in computer and forgets to lock the computer. If the current user is doing something else, there is no problem since the proper authentication data will not be provided by the user. However, if the user is nearby and typing on another device, the WACA will be vulnerable to attack since it will authenticate the user and keep the computer unlocked as far as the proper typing data is provided. 


In the second case, where there is no smartwatch data but there is an interaction with the keyboard. This means that someone who is different from the logged-in user is using the computer or the logged-in user's smartwatch currently can not provide data. In this case, WACA currently can not differentiate if somebody else is using or the current user's smartwatch can not provide data. The security policy can be adjusted to immediately logout even though it will decrease the convenient of the user, but this is a rare case as it can only occur if the battery dies or the smartwatch is broken. 

\noindent \textbf{Factors that may affect the performance.}
WACA is primarily designed for the organizations or offices where the passwords of the common computers may possibly be shared. However, the deployment of WACA is not restricted to those cases. It could also potentially be employed in applications where remote access is needed (e.g., web-based applications). The user may want to use WACA with his/her personal computer, where the user carries the computer wherever s/he goes. The high variation in the keyboard features like orientation, sizes, and types, or the current mood or the age of the user or the familiarity of the text may affect the efficiency of WACA. Therefore, for this type of personal usages, calibration may be needed by considering the trade-off between FAR and FRR. \revision{In this kind of personal usage scenarios, since the threat vectors will be more limited, the threshold can be extended to provide a more usable system.} We leave these as future work.

%% file: related_work.tex
\section{Related Work}\label{sec:related} In the literature, a number of works have been proposed for the use of biometrics in continuous user authentication~\cite{carrillo2003continuous,kang2006multi,azzini2008fuzzy,azzini2008impostor,altinok2003temporal,sim2007continuous,niinuma2010soft}. However, one of the desired features in the continuous authentication is \textit{transparency}. Hard biometrics like iris pattern or DNA are not applicable since they can not be extracted transparently. In another work~\cite{kwang2009usability}, a special mouse with a fingerprint sensor is proposed. In addition to requiring a custom mouse, its reliability is also an issue. The ease of counterfeiting fingerprints was shown and the fingerprint-based biometrics was easily bypassed~\cite{finger1,finger2}. Facial recognition methods may seem a good candidate; however, the liveliness detection is still an issue to be addressed and several attacks are possible under practical conditions~\cite{duc2009your,boehm2013safe}. In addition, several other biometrics like pulse-response~\cite{rasmussen2014authentication} or eye movements~\cite{eberz2015preventing} are also proposed. However, since these approaches require special equipments, deployment costs are increasing significantly.   

\noindent \textbf{Recent suspicions on keystroke dynamics.} Among all the biometrics, the most promising results are proposed using keystroke dynamics and mouse movements~\cite{banerjee2012biometric,teh2013survey}. However, in a recent work~\cite{tey2013can}, the reliability of classical keystroke dynamics are analyzed and an interface, called Mimesis, was  designed so that a user can mimic the typing rhythm of another user by using the feedback provided by Mimesis. In another study~\cite{serwadda2013examining}, the statistical attacks with bots generating synthetic typing patterns are examined for the conventional keystrokes biometrics. In our work, we test WACA against both these imitation and statistical attacks using similar configurations presented in these papers. We show that WACA is secure against the powerful imitation and statistical attacks with its design. Particularly, using 6 dimensions independently makes it harder to mimic and its high dimensional feature vector makes it harder to defeat using forged data. The detailed analysis of these attacks are given in Section~\ref{sec:attacks}.

\noindent \textbf{An inspirational work.} The closest work to our approach is proposed in~\cite{mare2014zebra} called ZEBRA.  
In ZEBRA, users are classified according to the sequence of interactions (e.g., typing, scrolling), where the user wears a bracelet with motion sensors and radio. \revision{ZEBRA has been shown as insecure in another work~\cite{huhta2015pitfalls}. Our work differs from ZEBRA in several ways. For the details, please see the Appendix.}

\noindent \textbf{Comparative evaluation of WACA.} In the literature, there is not widely accepted standard framework to compare device authenticators. However, Usability-Deployability-Security (UDS) framework proposed in~\cite{bonneau2012quest} is a highly accepted framework for web authentication schemes. To compare our work with its alternatives, we remove some of the irrelevant and non-applicable benefits and use only the relevant ones of UDS framework. The complete list of benefits can be found in~\cite{bonneau2012quest}. After also adding three new benefits, we end up with 18 benefits in total. Table~\ref{compare_eval} rates WACA using these 18 benefits. For space, we cannot compare WACA to all continuous authentication methods proposed in the literature. Therefore, we choose representatives for each continuous authentication method.

\begin{table}[!t]
\centering
\caption{Comparative evaluation of WACA using the UDS framework~\cite{bonneau2012quest} with continuous authentication alternatives. \label{compare_eval}}
\resizebox{.47\textwidth}{!}{

\setlength{\tabcolsep}{2pt} 
\renewcommand{\arraystretch}{1.5} 

\setlength{\columnseprule}{1pt}
\centering

\begin{tabular}{lccccccCcCccccccccc}
 & \rot{\textit{\normalsize Memorywise-Effortless}} & \rot{\textit{\normalsize Nothing-to-Carry}} & \rot{\textit{ \normalsize Physically-Effortless}} & \rot{\textit{ \normalsize Infrequent-Errors}} & \rot{\textit{ \normalsize Easy-Recovery-from-Loss}} & \rot{\textit{ \normalsize No-Constraint-on-Using-the-Device}} & \rot{\textit{ \normalsize Accessible}} & \rot{\textit{ \normalsize Negligible-Cost-per-User}} & \rot{\textit{\normalsize Resilient-to-Physical-Observation}} & \rot{\textit{ \normalsize Resilient-to-Targeted-Impersonation}} & \rot{\textit{\normalsize Resilient-to-Internal-Observation}} & \rot{\textit{\normalsize Resilient-to-Leaks-from-Other-Verifiers}} & \rot{\textit{\normalsize Resilient-to-Phishing}} & \rot{\textit{\normalsize Resilient-to-Theft}} & \rot{\textit{\normalsize Requiring-Explicit-Consent}} & \rot{\textit{\normalsize Unlinkable}} & \rot{\textit{\normalsize Insider-Identification}} & \rot{\textit{\normalsize Resilient-to-Insider-Threat}}   \\  \cmidrule{2-7} \cmidrule{8-9} \cmidrule{10-19}
 & \multicolumn{6}{|c|}{\Large Usability} &  \multicolumn{2}{c}{\Large Dep.} & \multicolumn{10}{c}{\Large Security}   \\ \midrule
{\normalsize Password} &  & \CIRCLE  &  & \CIRCLE  & \CIRCLE  & \CIRCLE  & \CIRCLE  & \CIRCLE  &  &  &  &  &  &  & \CIRCLE  & \CIRCLE  &  &   \\
{\normalsize Time-out} & \CIRCLE  & \CIRCLE  & \LEFTcircle & \CIRCLE  & \CIRCLE  & \CIRCLE  & \CIRCLE  & \CIRCLE  &  & na & \CIRCLE  &  & \CIRCLE  & na &  & na &  &   \\
{\normalsize Proximity~\cite{landwehr1997protecting,corner2002zero}} & \CIRCLE  & \LEFTcircle & \LEFTcircle & \LEFTcircle & \LEFTcircle & \CIRCLE  & \CIRCLE  & \LEFTcircle & \LEFTcircle & na & \CIRCLE  &  &  &  &  &  &  &   \\
{\normalsize Face~\cite{beunder2014design,niinuma2010continuous}} & \CIRCLE  & \CIRCLE  & \LEFTcircle &  &  & \LEFTcircle & \LEFTcircle & \LEFTcircle & \CIRCLE  & \LEFTcircle &  & \CIRCLE  &  & \CIRCLE  & \LEFTcircle & \CIRCLE  & \CIRCLE  & \CIRCLE  \\
{\normalsize Fingerprint~\cite{kwang2009usability} }& \CIRCLE  & \CIRCLE  & \LEFTcircle &  &  &  & \LEFTcircle &  & \CIRCLE  & \LEFTcircle &  & \CIRCLE  &  & \CIRCLE  & \LEFTcircle & \CIRCLE  & \CIRCLE  & \CIRCLE  \\
{\normalsize Eye-movement~\cite{eberz2015preventing}} & \CIRCLE  & \CIRCLE  & \LEFTcircle &  &  &  & \LEFTcircle &  & \CIRCLE  & \CIRCLE  & \CIRCLE  & \CIRCLE  & \CIRCLE  & \CIRCLE  & \CIRCLE  & \CIRCLE  & \CIRCLE  & \CIRCLE   \\
{\normalsize Keystroke~\cite{monrose1997authentication}} & \CIRCLE  & \CIRCLE  & \LEFTcircle & \LEFTcircle &  & \CIRCLE  & \CIRCLE  & \CIRCLE  & \LEFTcircle &  & \CIRCLE  &  & \CIRCLE  &  & \CIRCLE  &  &  & \CIRCLE  \\
{\normalsize ZEBRA~\cite{mare2014zebra}} & \CIRCLE  & \LEFTcircle & \LEFTcircle & \LEFTcircle & \LEFTcircle & \CIRCLE  & \CIRCLE  & \LEFTcircle &  &  & \CIRCLE  & \CIRCLE  & \CIRCLE  & \LEFTcircle & \CIRCLE  & \CIRCLE  & \LEFTcircle & \CIRCLE    \\
{\normalsize \textbf{WACA (this work)}} & \CIRCLE  & \LEFTcircle & \LEFTcircle & \LEFTcircle & \LEFTcircle & \CIRCLE  & \CIRCLE  & \LEFTcircle & \CIRCLE  & \CIRCLE  & \CIRCLE  & \CIRCLE  & \CIRCLE  & \CIRCLE  & \CIRCLE  & \CIRCLE  & \CIRCLE  & \CIRCLE \\
\bottomrule
\multicolumn{19}{l}{\small {\normalsize \CIRCLE} $=$ offers the benefit; {\normalsize \LEFTcircle} $=$ almost offers the benefit; no circle $=$ does not offer the benefit.} \\

\end{tabular}}


\end{table}

WACA captures the sensor readings through a smartwatch without interrupting the user, i.e., unobtrusively. However, unlike time-out or classical keystroke dynamics, it requires an extra channel to collect data, but obviously a smartwatch is a not a customized hardware, i.e., it is an off-the-shelf device, so we say it partially supports the benefit of \textit{Nothing-to-Carry} and since its error is very low, it also offers the benefit of \textit{Infrequent-Errors}. On the other hand, WACA outperforms all other methods in terms of security benefits. In addition to WACA, eye-movement based authentication method also seems as secure as WACA. However, WACA's performance for usability and deployability is better. For example, WACA offers much lower error rates and eye-movement based methods require a special eye or gaze-trackers and the user should be in a certain distance and in front of the eye tracker which obstructs the usability of the eye-movement based methods. They are more convenient for challenge-response type authentication methods~\cite{sluganovic2016using} even though they have the capability to provide data continuously and transparently. In brief, our conclusion from this comparative evaluation shows that WACA offers better security benefits while keeping the usability at the same level as other notable methods.

%% file: conclusion.tex
\section{Conclusion} \label{sec:conc}

Wearables such as smartwatches and 
fitness trackers 
carried by individuals have grown exponentially in a short period of time. 
It is estimated that one in four smartphone owners will also be using a wearable device such as a smartwatch. This ubiquity of wearable devices make them a perfect candidate to utilize in continuous authentication settings. The continuous authentication allows users to be re-verified periodically throughout the login session without breaking the continuity of the session. In this paper, we introduced a novel Wearable-Assisted Continuous Authentication (WACA) utilizing the sensory data from the built-in motion sensors available on smartwatches. WACA is a practical and usable wearable-assisted continuous authentication system that 
combines the functionality of wearables and usability of continuous authentication.  
Particularly, WACA decreases the vulnerable time window of a continuous authentication system to as low as 20 seconds, prevents the privilege abuse and insider attacks 
and also allows the insider threat identification. Moreover, we evaluated the efficacy and robustness of WACA with real data from real experiments. 
The results showed that WACA could achieve 1\% EER for 30 seconds or $2-3\%$ EER for 20 seconds of data collection time and error rates are as low as $1\%$ with almost a perfect (99.2\%) insider threat identification rate. Furthermore, WACA was tested and shown as secure against different powerful adversaries (imitation and statistical attacks) and achieved a minimal overhead on the utilization of the system's resources. As a future work, it would be interesting to test WACA to deploy in 
authentication settings 
combining keystroke dynamics from both keyboard and smartwatch sensors.



%% file: appendix.tex

\setcounter{figure}{0}
\setcounter{equation}{0}

\section{Model training time for the insider threat detection} \label{sec:timings}
In Table~\ref{iden_diff3} and \ref{iden_diff4}, we present the times to built the model for MLP algorithm, which is used for the insider threat detection in Section~\ref{sec:eval}.
\begin{table}[!ht]
    \centering
    
    \caption{Scenario 1 in Section~\ref{sec:eval}. \label{iden_diff3}}
      \resizebox{.8\columnwidth}{!}{%
    \begin{tabular}{cc}
    \begin{tabular}{cccccc}
     \multicolumn{6}{c}{\centering \textbf{Time taken to build the model (seconds)}} \\ \hline \hline \bigstrut
     & \multicolumn{4}{c}{\centering Training Set} &  \\  \cline{3-4} \bigstrut
     Sample size   & 1     & 2     & 3     & 4     & 5     \\ \hline  \bigstrut
   
    1500  & 0.99  & 2.02  & 3     & 3.99  & 4.94  \\ \bigstrut
    1000  & 1     & 1.99  & 2.98  & 3.95  & 4.94  \\ \bigstrut
    500   &  0.98  & 2.02  & 3.01  & 3.98  & 4.95  \\ \bigstrut
    250   & 1     & 1.95  & 2.93  & 3.92  & 4.95 \\ \bottomrule \bigstrut
    
    \end{tabular}
    \end{tabular}}
 \end{table}
  

 \begin{table}[!ht]
    \centering
    
    \caption{Scenario 2 in Section~\ref{sec:eval}. \label{iden_diff4}}
      \resizebox{.8\columnwidth}{!}{%
    \begin{tabular}{cc}
    \begin{tabular}{cccccc}
     \multicolumn{6}{c}{\centering \textbf{Time taken to build the model (seconds)}} \\ \hline \hline \bigstrut
     & \multicolumn{4}{c}{\centering Training Set} &  \\  \cline{3-4} \bigstrut
     Sample size   & 1     & 2     & 3     & 4     & 5     \\ \hline  \bigstrut
   
    1500  & 1     & 1.97  & 2.97  & 4.01  & 4.99  \\ \bigstrut
    1000  & 1     & 1.99  & 2.95  & 4.04  & 4.93  \\ \bigstrut
    500   & 0.99  & 2     & 2.98  & 3.98  & 4.98  \\ \bigstrut
    250   & 0.99  & 1.96  & 3.02  & 3.96  & 4.96\\ \bottomrule \bigstrut
    
    \end{tabular}
    \end{tabular}}
 \end{table}


 \section{Statistical attack algorithm}  \label{sec:stat}
 In Algorithm~\ref{alg:stat}, we present the pseudo-code of the attack algorithm given in~\ref{subsec:attack2}.
 \begin{algorithm}[!ht] 
\caption{Calculate EER for a statistical attacker. 
\label{alg:stat}}
\begin{algorithmic}[1] 

 \REQUIRE $Samples_{MXN}$[]:  M is \# of samples and N is \# of features
 \REQUIRE $outNumber$: \# of generated forged samples
 \REQUIRE $binNumber$: \# of bins 
 \REQUIRE $topBins$: \# of top bins used to generate forge samples
 \ENSURE $new\_eer$: \# new error rate against the attack   
   \FOR{each  $user$} 
   \STATE{$victim \leftarrow user$;}
   \STATE{$victimSamples \gets$ getSamples($victim$); } 
   \STATE{$attackSamples \gets$ getSamples($\sim victim$)};
   \STATE{ $combin$[] $\gets$ ComGen($N, outNumber, topBins$);}
    \FOR{each forgeid $s_i  \in  attackSamples$} 
    \FOR{each feature $f_j \in attackSamples$} 
      \STATE{[$freq,edges$] $\gets$ histGen();} 
      \STATE{[$\sim$,$index$] $\gets$ sortBins($freq$);} 
      \STATE{$index(topBins+1:end) \gets$ [];}
      \STATE{$m \gets edges(index(combin[f_j,forgeid]))$;}
      \STATE{$forgedSamples \gets$ random([$m$, $m+1$]);}
    \ENDFOR
    \ENDFOR
     \STATE{$victimSamples \gets$ addSamples($forgedSamples$);}
     \STATE{$D \leftarrow$ calculateDissMatrix($TestingSamples$);}
     \STATE{$eer\_for\_victim \leftarrow$ calculateEER($D$);}
    \ENDFOR\\  
    \STATE{$new\_eer \leftarrow$ mean($eer\_for\_victim$);}
   \RETURN{$new\_eer$}       
\end{algorithmic}
\end{algorithm}
 

\section{How it differs from ZEBRA.}  \label{sec:zebra}

\revision{Serious design flaws has been detected. In ZEBRA, the attacker has control on the data that authenticator components receives. Therefore, an attacker can opportunistically choose a subset of the sequences to imitate, instead of na\"ively trying to imitate every activity the user is doing. Although the idea of using a wearable in continuous authentication settings seems similar with ours, our design differs from ZEBRA in many ways to tackle those flaws and strengthen our design.
First and most importantly, instead of sequence of interactions, we use a variant of keystroke dynamics as an authentication factor, which was extensively well-studied in the research literature and shown as a unique behavioral biometric for each person (see~\cite{ali2016keystroke} for a very recent survey). Behavioral biometrics are harder to imitate and the keystrokes are much shorter than the interactions, which makes it harder to be imitated by a nearby shoulder-surfing attacker.
We also note that we tested WACA against imitation attacks in Section~\ref{subsec:attack1}. Moreover, we also use each axis of the sensor data to create another feature while in ZEBRA, before using a sensor data, the magnitude of each sensor's data is calculated by the root mean square formula. Each axis of the sensor data is independent from each other. Therefore, an imitation attack~\cite{tey2013can} is possible only if one can mimic another one's typing behaviour in six different dimensions. This increases the strength of WACA significantly against imitation attacks. 
Not only these security measures, WACA has some other minor design details that makes it more usable. For example, while ZEBRA requires wearing the bracelet on the hand that the mouse is controlled, WACA does not have a limitation on which hand the smartwatch is worn. Finally, since WACA utilizes a behavioral biometric, it also allows insider threat identification, where ZEBRA is not tested for this purpose.}